\documentclass{PoS}
\usepackage{macros_static,mathrsfs,amssymb,amsmath,amsfonts,epsfig,cite,graphics}

\def\LALPHA{\hbox{\epsfxsize=2.0 true cm
                           \epsfbox{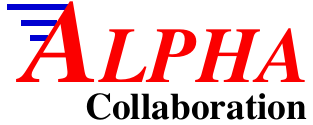}}
                   }
\newcommand{\onecol}[2]{
        \begin{minipage}[t]{#1}{#2\vfill} \end{minipage}
        }

\title{$B$ meson spectrum and decay constant from ${\rm N_{\rm f}}=2$ simulations}

\ShortTitle{}

\author{\LALPHA \hfill
        \onecol{4.0cm}{\vspace{-1.9cm}\it Edinburgh 2010/37 \\ MKPH-T-10-35
          \\ DESY 10-209  \\ SFB/CPP-10-116 \\MS-TP-10-26 \\SHEP-1041
          \vspace{.5cm}
        }} 

\author{\speaker{Beno\^\i t Blossier} $^a$, John Bulava$^b$, 
Michele Della Morte $^c$,  Michael Donnellan $^b$, 
Patrick Fritzsch $^d$, \speaker{Nicolas Garron} $^e$,  
Jochen Heitger $^f$, Georg von Hippel $^c$,
Bj\"orn~Leder $^g$, 
Hubert Simma $^b$, Rainer Sommer $^b$ \vspace{0.4cm} \\
\llap{$^a$} Laboratoire de Physique Th\'eorique, CNRS et Universit\'e Paris-Sud XI, 
B\^atiment 210, 91405~Orsay~Cedex, France\\
\llap{$^b$} NIC, DESY, Platanenallee 6, 15738 Zeuthen, Germany \\
\llap{$^c$} Universit\"at~Mainz, Institut~f\"ur~Kernphysik, Becherweg~45, 55099~Mainz, Germany \\
\llap{$^d$} School of Physics and Astronomy, University of Southampton, Southampton, SO17 1BJ, U.K. \\
\llap{$^e$} School of Physics and Astronomy, University of Edinburgh, Edinburgh EH9 3JZ, U.K.\\
\llap{$^f$} Universit\"at M\"unster, Institut f\"ur Theoretische Physik,
Wilhelm-Klemm-Strasse 9, 48149 M\"unster, Germany \\
\llap{$^g$} Universit\"at Wuppertal, Gaussstr. 20, 42119 Wuppertal, Germany \\
}

\abstract{We report on the status of an ALPHA Collaboration project to extract
quantities for $B$ physics phenomenology from ${\rm N_{f}}=2$ lattice simulations. 
The framework is Heavy Quark Effective Theory (HQET) expanded up to the first order 
of the inverse $b$-quark mass. The couplings of the effective theory are determined 
by imposing matching conditions of observables 
computed in HQET with their counterpart computed in QCD. That program, based on ${\rm N_{f}}=2$ simulations 
in a small physical volume with \SF boundary conditions, is now almost finished. 
On the other side the analysis of configurations selected from the CLS ensembles, 
in order to measure HQET hadronic matrix elements, 
has just started recently so that only results obtained
at a single lattice spacing, $a=0.07\,\fm$, 
will be discussed. We give our first results for the $b$-quark mass
and for the $B$ meson decay constant.}

\FullConference{The XXVIII International Symposium on Lattice Field Theory, Lattice2010\\
		June 14-19, 2010\\
		Villasimius, Italy}

\begin{document}

\section{Introduction}

Now that the amount of available experimental data on
        beauty physics coming from the $B$ factories and the Tevatron is
        large, and even more data are expected from the LHC, precision
        tests of the Standard Model and searches for New Physics have
        become possible in this area of flavor physics. Unfortunately, the
        theoretical uncertainty, mainly because of the difficult to estimate
        long-distance effects due to confinement, is currently limiting the
        impact
of future experimental measurements on New 
Physics models. Lattice QCD makes it possible to reach a few percents of theoretical error on those non-perturbative 
hadronic contributions, but care is needed to obtain reliable results for $b$-quark physics. Indeed one has 
to keep under control simultaneously the finite size effects and, particularly, the discretisation effects (the 
lattice spacing should  be smaller than the Compton wavelength of the $b$-quark) induced by the simulation. In 
practice it is not possible to control both effects in one simulation. Different approaches have 
been proposed in the literature (see for example~\cite{Heitgerlat10} for a recent review).
The ALPHA Collaboration has followed a strategy discussed in detail in \cite{HeitgerNJ} - 
\cite{BlossierJK}: it is based on the use of HQET, in which the hard degrees of freedom $\sim\, m_{\rm b}$ 
are integrated out and taken into account by an expansion in the inverse $b$-quark mass $m_{\rm b}$ . 
As discussed in those papers and also in earlier work, the benefit is the suppression of
large discretisation effects which may arise in hadronic quantities when the theory is 
regularised on the lattice.
The difficult aspect of that method is that a matching with QCD, which 
is the field theory believed to describe the strong interactions, is needed to absorb ultraviolet divergences appearing 
in the effective theory. In HQET those come as inverse powers of the lattice spacing and thus have to be removed non-perturbatively before the continuum limit can be taken. 
This method has been tested in the quenched approximation to study 
the $B_{\rm s}$ meson spectrum 
\cite{BlossierVZ} and to determine the $b$-quark mass \cite{DellaMorteCB}, the decay constant
$f_{\rm B_{\rm s}}$ \cite{BlossierMK}, and the coupling $g_{{\rm B}^{*}{\rm B}\pi}$ \cite{MDonnellan}.

\begin{figure}[b]
\begin{center}
\includegraphics*[width=12cm, height=8cm]{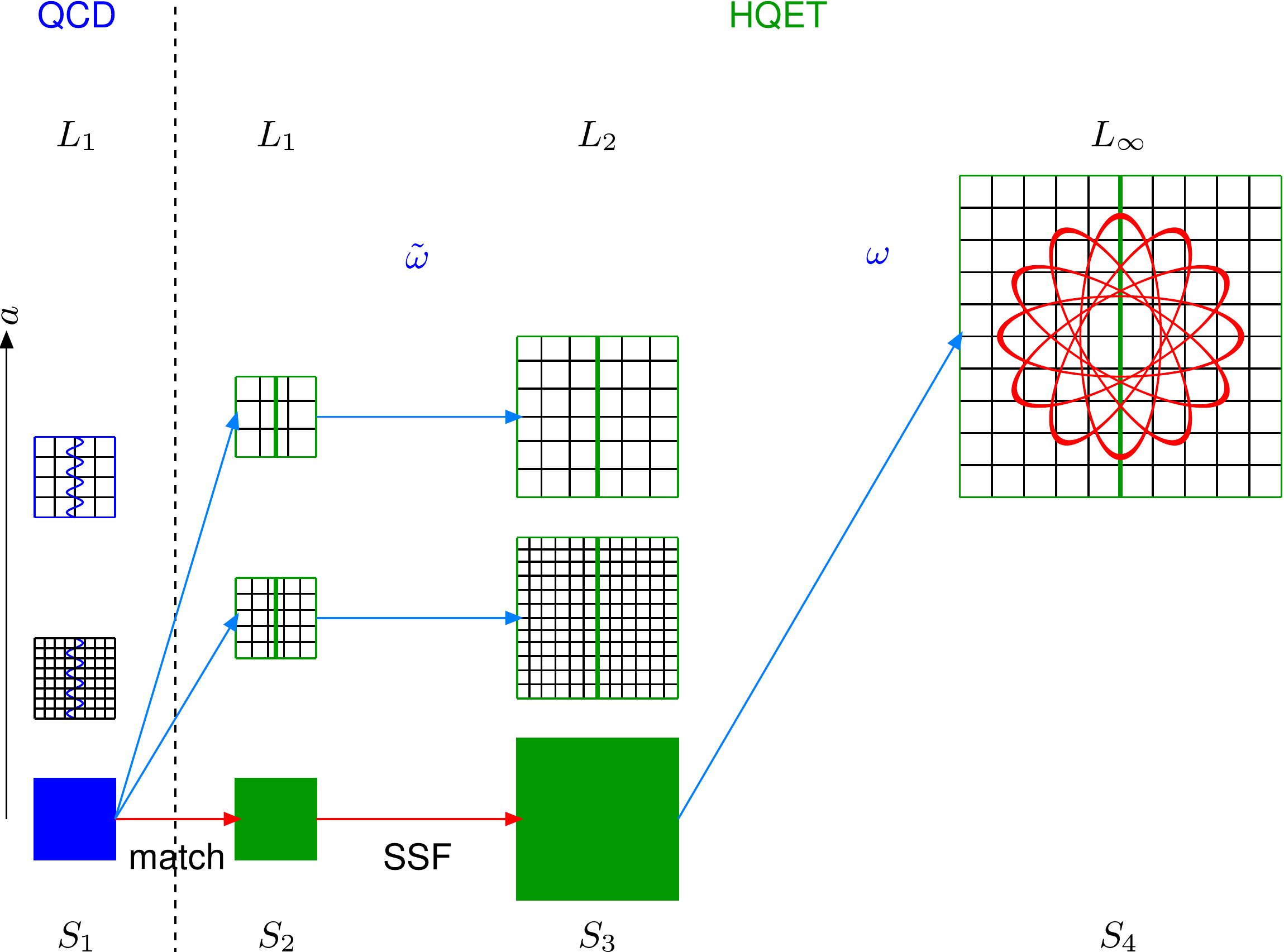}
\end{center}
\caption{Sketch of the strategy followed by the ALPHA Collaboration to compute $B$ 
physics observables on the lattice.}
\label{figstrategy}
\end{figure}

The strategy, sketched in \fig{figstrategy}, is now being extended to the more realistic ${\rm N_{f}}=2$ situation. 
In the next section, more details are given about the determination of HQET couplings, performed in a small physical volume (of space extent $\sim$ 0.5 fm). In section 3, we report on the analysis of a subset of ensembles 
generated within the CLS effort, to get HQET energies and matrix elements. Section 4 contains our conclusions.

\section{Computation of the relevant HQET parameters}

We write the HQET Lagrangian as
\bes
\lag{HQET}(x) &=&  \lag{stat}(x) - \omegakin\Okin(x)
        - \omegaspin\Ospin(x)  \,,
\ees
where the lowest order (static) term is
\be
\label{e:statact}
\lag{stat}(x) =
\heavyb(x) \,D_0\, \heavy(x) \;,
\ee
and the first order corrections in $1/m_{\rm b}$ are 
\bes
  \Okin(x) &=& \heavyb(x){\bf D}^2\heavy(x) \,,\quad
  \Ospin(x) = \heavyb(x){\boldsymbol\sigma}\!\cdot\!{\bf B}\heavy(x)\,.
\ees
We are also interested in the time component of the heavy-light axial current $A_0$.
Considering only the terms which contribute to zero-momentum correlation functions,
we write it as
\bes
 \label{e:ahqet}
 \Ahqet(x)&=& \zahqet\,[\Astat(x)+  \cah{1}\Ah{1}(x)]\,, \\
 \Ah{1}(x) &=& \lightb(x){1\over2}
            \gamma_5\gamma_i(\nabsym{i}-\lnabsym{i}\;)\heavy(x)\,,
 \quad \Astat(x) = \lightb(x)\gamma_0\gamma_5\heavy(x)\,,
 \label{e:dahqet}
\ees
where $\nabsym{i}$ denotes the symmetric derivative.

In this section we present the computation of 
$\omegakin, \omegaspin, \zahqet,  \cah{1}$ and $\mhbare$ 
(the energy shift which in the static theory absorbs the $1/a$ 
divergence of the static energy and at order $\minv$ absorbs a $1/a^2$ term).
These parameters can be used for a computation of the $b$-quark mass, the  heavy-light meson
decay constants ($\fB$ or $\fBs$), as well as for a determination of 
the spectrum of heavy-light mesons, including the hyperfine mass splitting.
We follow the strategy presented in~\cite{BlossierJK},
where the computation was done in the quenched approximation.
We recall here the essential ingredients and refer the reader 
to this paper for any unexplained notation and for more detailed explanations.
This computation is done non-perturbatively (in the strong coupling), 
and at the $1/\mb$ order of the heavy quark expansion. 
The light quarks are simulated with 2-flavor Clover-improved Wilson fermions, 
and for the discretisation of the heavy quark we use the so-called HYP1 and HYP2
actions~\cite{DellaMorte:2005yc}. More details about the implementation 
can be found in~\cite{DellaMorte:2007qw}.
The simulations considered in this section use the Schr\"odinger 
functional setup. 

\begin{figure}[htb]
\begin{center}
\begin{tabular}{cc}
\includegraphics*[width=6cm, height=5cm]{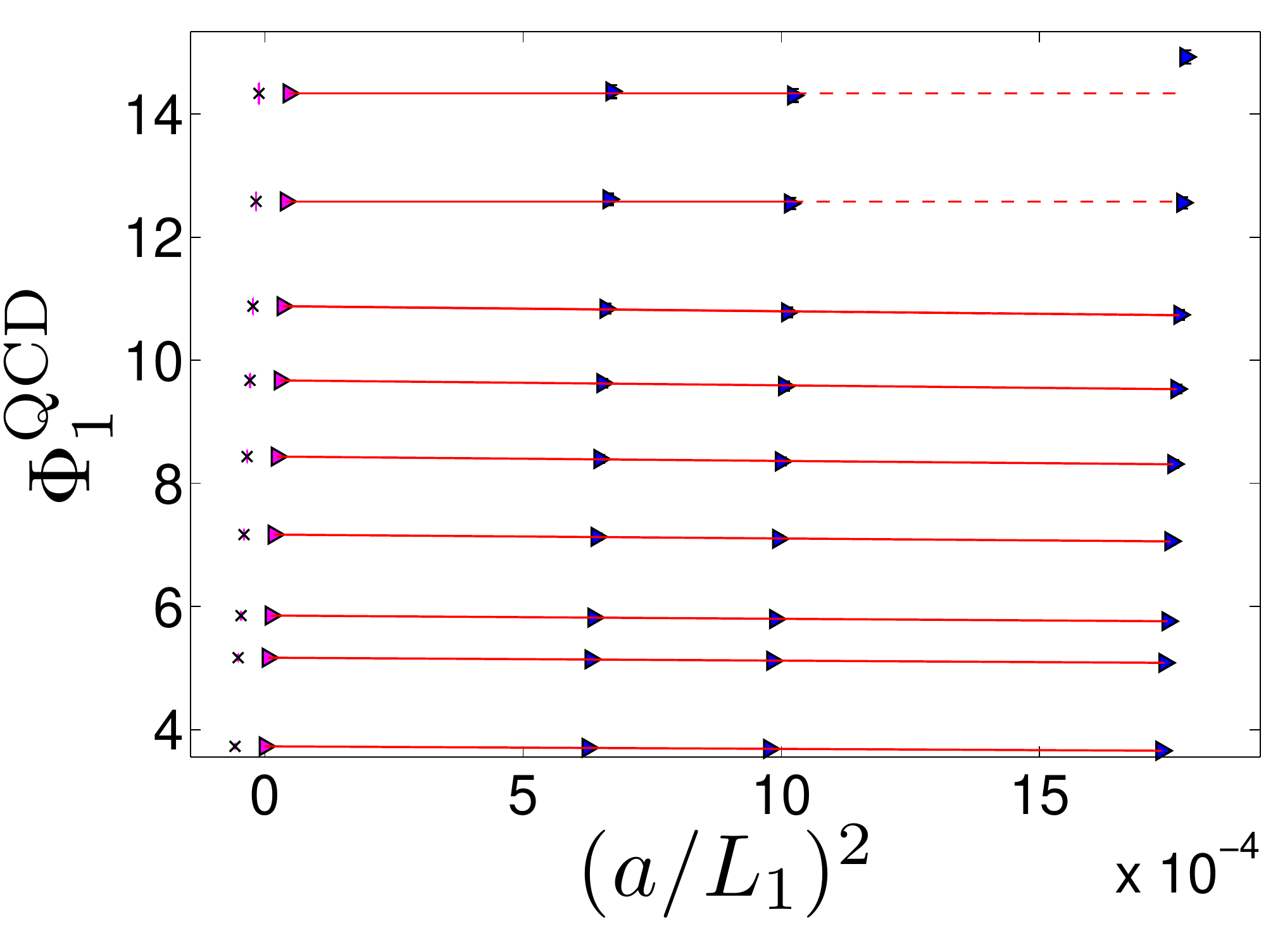}
&
\includegraphics*[width=6cm, height=5cm]{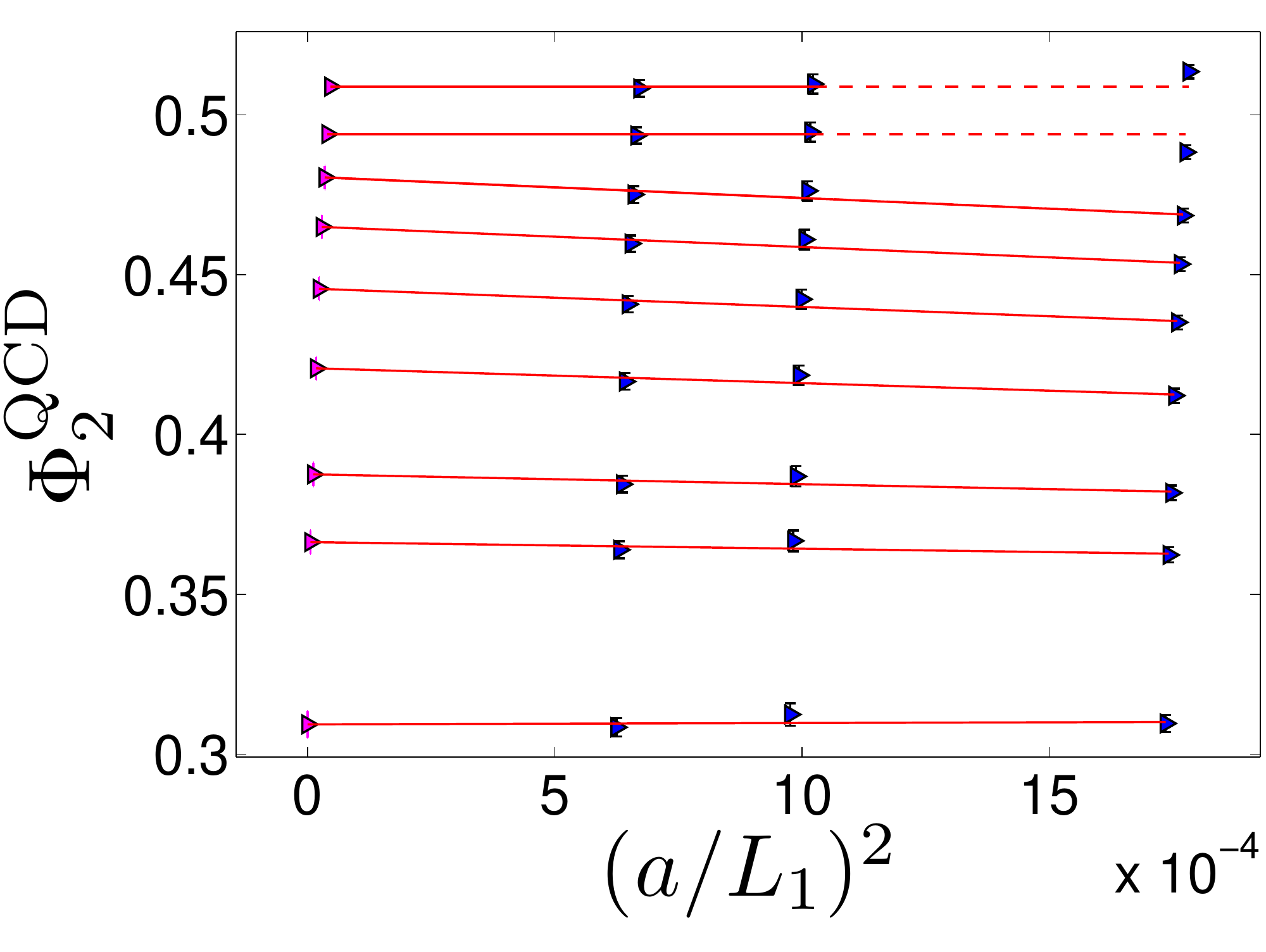}
\end{tabular}
\end{center}
\vspace{-0.5cm}
\caption[]{Continuum extrapolation of the QCD observables $\Phi_1$ and $\Phi_2$,
which are proportional to the finite volume meson mass and to the finite volume 
logarithm of the decay constant, respectively.
For $\Phi_1$ we have included an error in the continuum coming from the
renormalisation of the quark mass (cross on the left).
Results are shown for the nine different values of the heavy quark mass.
}
\vspace{0.5cm}
\label{figPhi12_L1}
\end{figure}

Following~\cite{BlossierJK}, we define a set of observables $\Phi_{i=1,\ldots, 5}$
that we collect in a vector $\Phi$. 
In the continuum and large volume limits, $\Phi_1$ is proportional to 
the meson mass and $\Phi_2$ to the logarithm of the decay constant, respectively.
$\Phi_3$ is used to determine the counter-term of the axial current, 
and $\Phi_{4,5}$ for the determination of the kinetic and magnetic term,
respectively. 
We first consider a small volume (of linear space extent $L_1\sim 0.5$ fm),
where the $b$-quark mass can be simulated with discretisation effects under control.
In this volume, we compute the five  observables $\Phi_i$ in QCD, for four different lattice spacings.
In this set of simulations, the light quark masses are set to 0, 
while for the (RGI) heavy quark mass $M$ we have chosen nine different values
such that $z=L_1 M= 4,6,7,9,11,13,15,18,21 $. The lightest masses
correspond approximately to the charm and the heaviest to the bottom
quark. In this work we will focus on the heaviest ones.
The continuum limit of each observable is obtained by a linear extrapolation in $(a/L_1)^2$
of the three finest lattice spacings (except for the two heaviest masses
where only the two finest lattices are used):
$\Phi^{\rm QCD}_i(L_1, M, 0) = \lim_{a\to 0} \Phi_i^{\rm QCD}(L_1,M,a)$.
This set of simulations is represented by $S_1$ in \fig{figstrategy}.
The continuum extrapolations of the first two observables are shown 
in \fig{figPhi12_L1}.

In another set of simulations called $S_2$, we compute the corresponding quantities 
in the effective theory, using the same value of the physical volume.
We then impose the QCD observables to be equal to their HQET expansion
at the $1/\mb$ order~\footnote{Terms of order $1/\mb^2$ are dropped 
without notice.}.
This matching can be written in the following way:
\begin{equation}
\label{eq:matchL1}
\Phi^{\rm QCD}_i(L_1,M,0) = \eta_i(L_1,a) + \sum_j \varphi_{ij}(L_1,a)\,\tilde\omega_j(M,a) \;,
\end{equation}
where $\eta$ and $\varphi$ are computed by lattice simulations 
for different values of the lattice spacing $a$. 
In other words the matching equations determine the set of parameters 
$\tilde\omega =  \varphi^{-1}\,[\Phi^{\rm QCD} -\eta]$.
For example, in the the static approximation $\varphi$ is diagonal 
and (up to a factor $L_1$) $\eta_1$ is given by the static energy, 
$\varphi_{11}$ is one and $\tilde\omega_1$ is the bare quark mass in static approximation.
In table \ref{table_hqet_param_def}, we list the various parameters 
together with their values 
in the static approximation and in the classical limit.
 \begin{table}[!htb] 
 \hspace{-1.cm} 
 \begin{center} 
 \begin{tabular}{|c|c|c|c|c|}
 \hline 
 $\omega_i$  & definition & classical value  &  static value \\[0.2ex]
 \hline 
 $\omega_1$  & $\mhbare$  &  $\mb$ & $ \mhbare^\stat$  \\ [+0.7ex]
 $\omega_2$  & $\log(Z_A^{\rm HQET})$  &  0 & $ \log(\zastat) $  \\ [+0.7ex] 
 $\omega_3$  & $ \cah{1}   $  &  $-1/(2\mb)$ & $ a\castat$  \\ [+0.7ex]
 $\omega_4$  & $\omegakin  $  &  $1/(2\mb)$ & $0$  \\ [+0.7ex]
 $\omega_5$  & $\omegaspin $  &  $1/(2\mb)$ & $0$  \\ [+0.7ex]
 \hline 
 \end{tabular} 
 \end{center} 
 \caption[ ]{\footnotesize Notation for HQET parameters, their values in classical 
   and static approximation. For the numerical value of $\castat$, we use the formula
   given in~\cite{Grimbach:2008uy} . }
 \label{table_hqet_param_def} 
 \end{table}

We then compute $\eta$ and $\varphi$ in a larger volume of space extent $L_2=2L_1$
using the same set of lattice spacings as the one used in the previous step.
This step is represented by the set $S_3$.
The observables in this volume are obtained from the parameters $\tilde \omega(M,a)$
determined in the previous step:
\begin{equation}
\label{eq:matchL2}
\Phi(L_2,M,0) = \lim_{a\to0} 
\left(\eta(L_2,a) + \varphi(L_2,a)\tilde\omega(M,a) 
\right)
\;.
\end{equation}
The continuum limit can be taken because the divergences
cancel out in the previous equation.
This procedure can then be re-iterated until the volume reached is large enough 
for finite size effects to be negligible,
 typically around $(2\, \fm)^3$. 
In practice, it turns out that three different volumes are enough 
($L_1,L_2$ and the large volume one). 
Thus, the HQET parameters that can be used in large volume simulations
(denoted as $S_4$) 
are given by
\begin{equation}
\omega(M,a) = \varphi^{-1}(L_2,a)\left[ \Phi(L_2,M,0) - \eta(L_2,a) \right] \;.
\end{equation}
Finally we perform a small interpolation in $\beta$ in order to obtain
the parameters at the lattice spacing used in the CLS ensembles.

\begin{figure}[htb]
\begin{center}
\begin{tabular}{cc}
\includegraphics*[width=6cm, height=5cm]{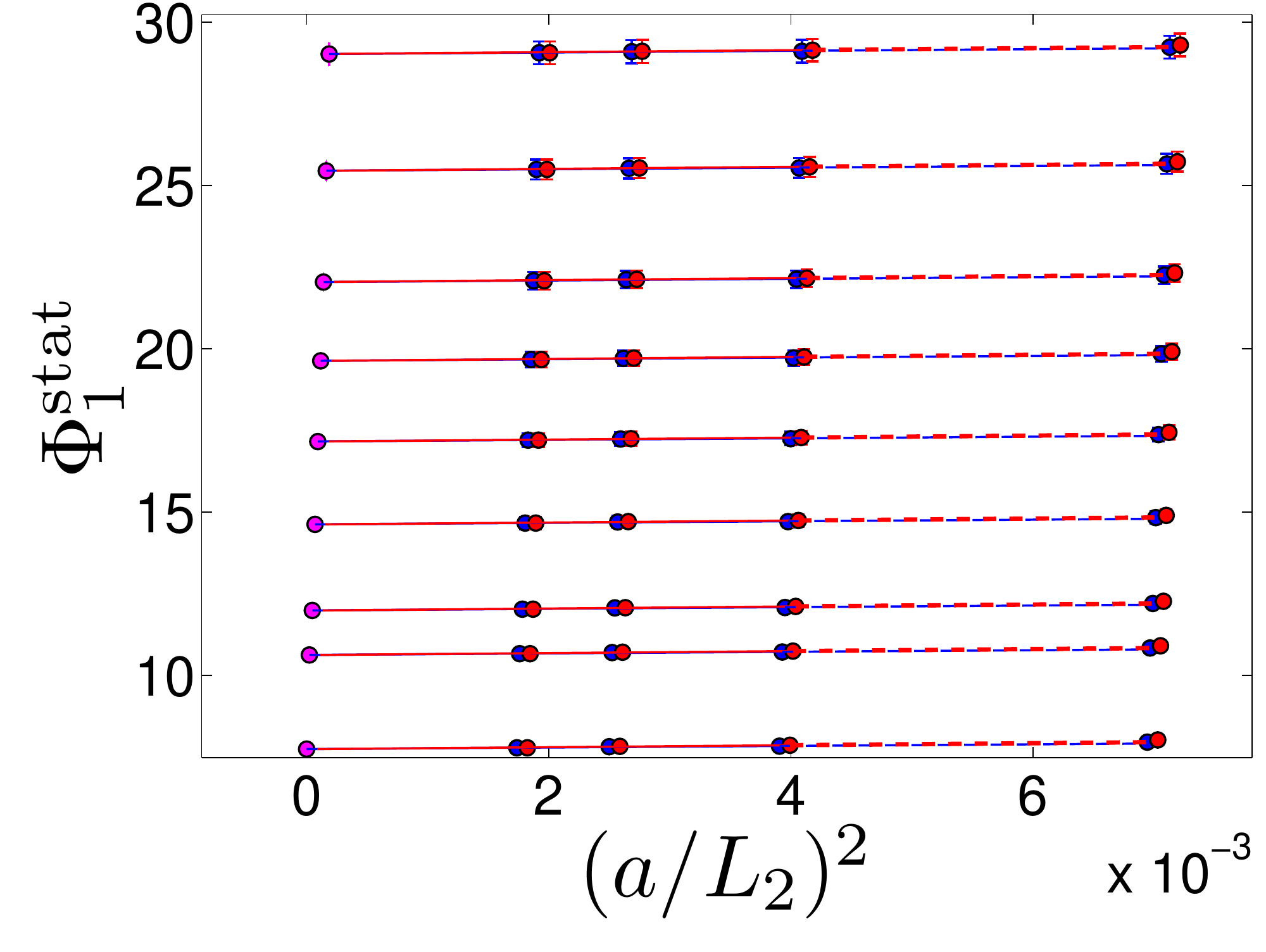}
&
\includegraphics*[width=6cm, height=5cm]{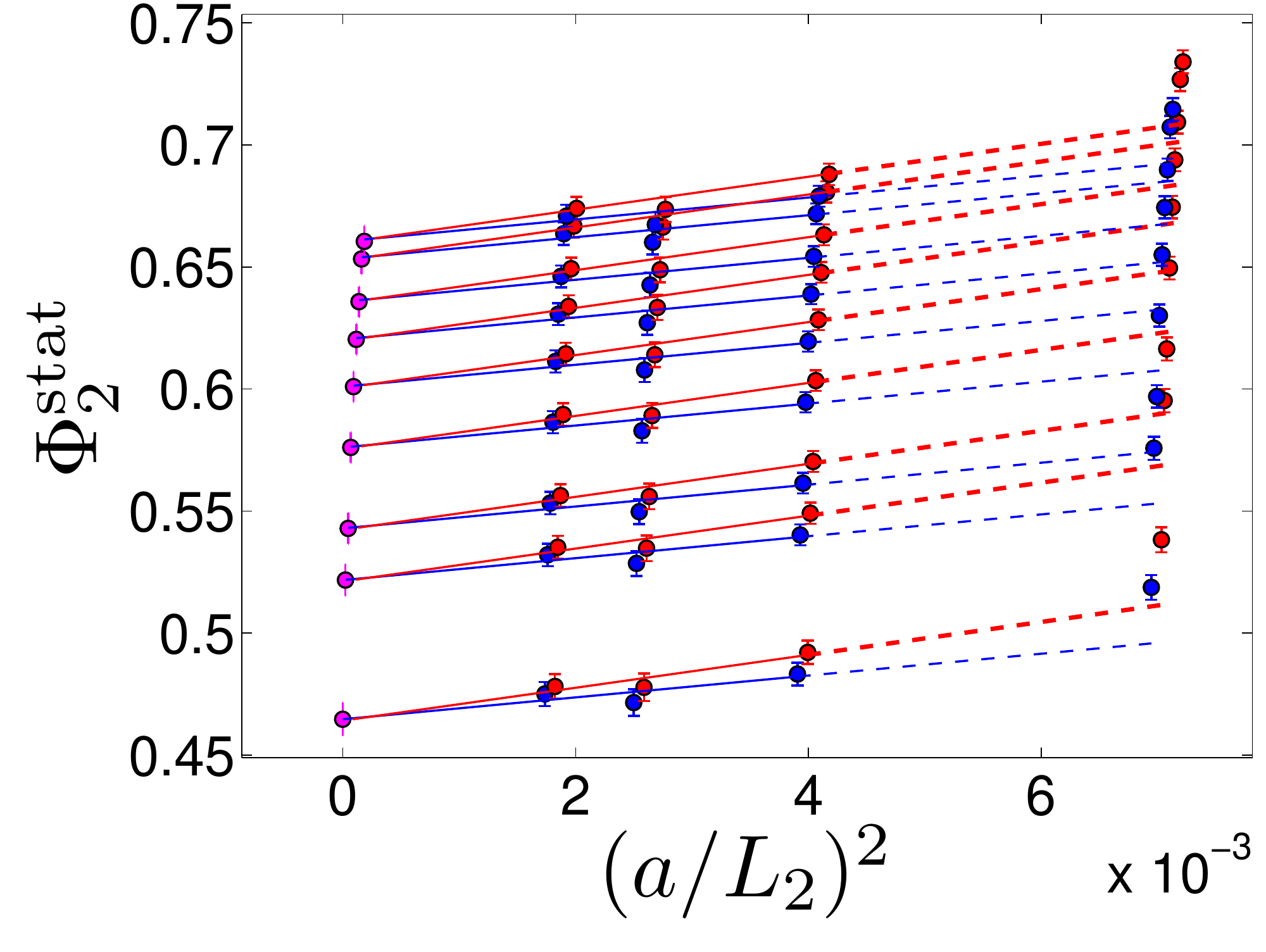}
\end{tabular}
\end{center}
\vspace{-0.5cm}
\caption[]{Continuum extrapolation of the static approximation of $\Phi_1$ and $\Phi_2$ 
in the volume of space extent $L_2$. The two different colors correspond to 
two different discretisations of the static propagator : red for HYP1 and blue 
for HYP2. Only the three finest lattice have been used in the continuum extrapolation.}
\vspace{0.5cm}
\label{figPhi12_stat_L2}
\end{figure}

\begin{figure}[htb]
\begin{center}
\begin{tabular}{cc}
\includegraphics*[width=6cm, height=5cm]{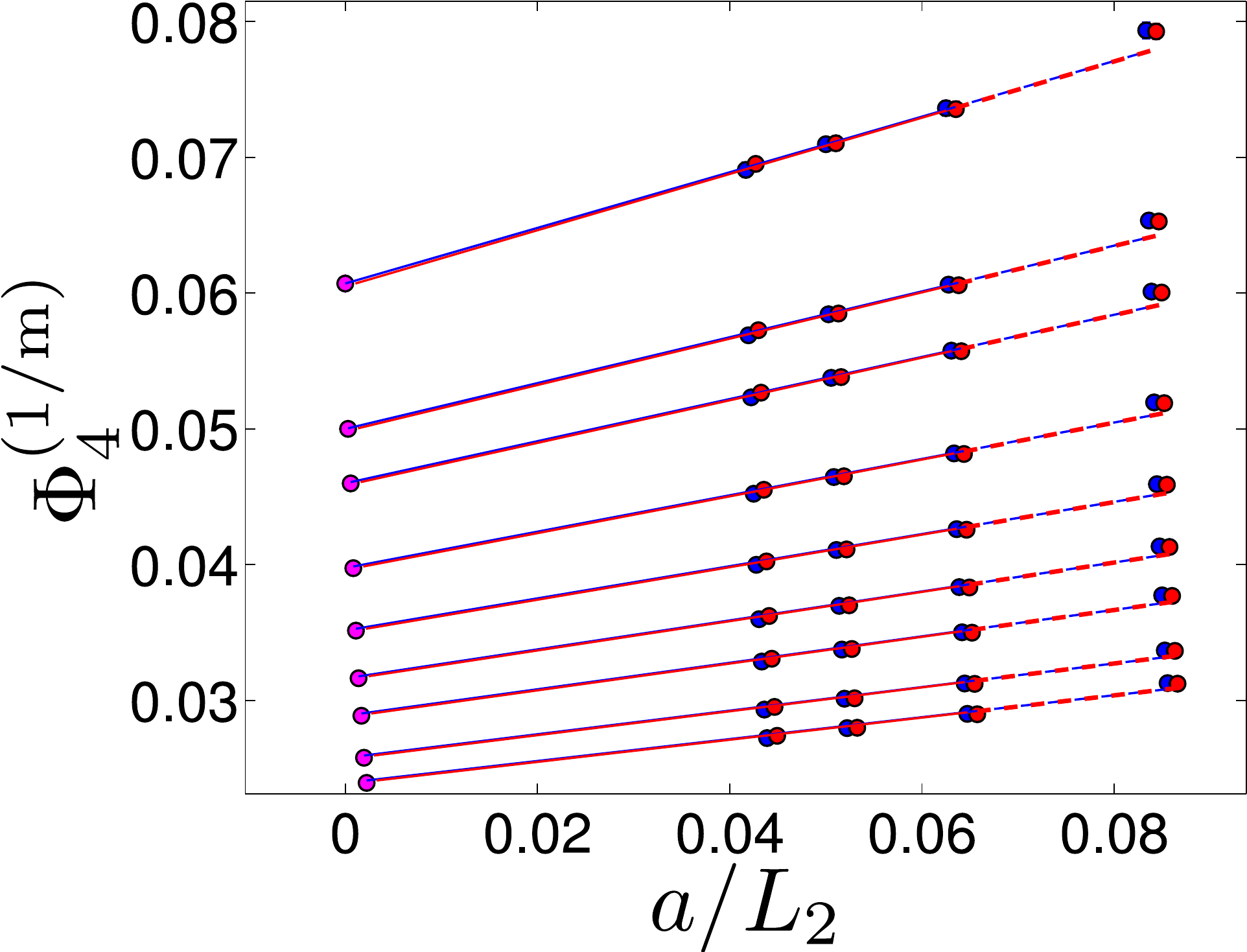}
&
\includegraphics*[width=6cm, height=5cm]{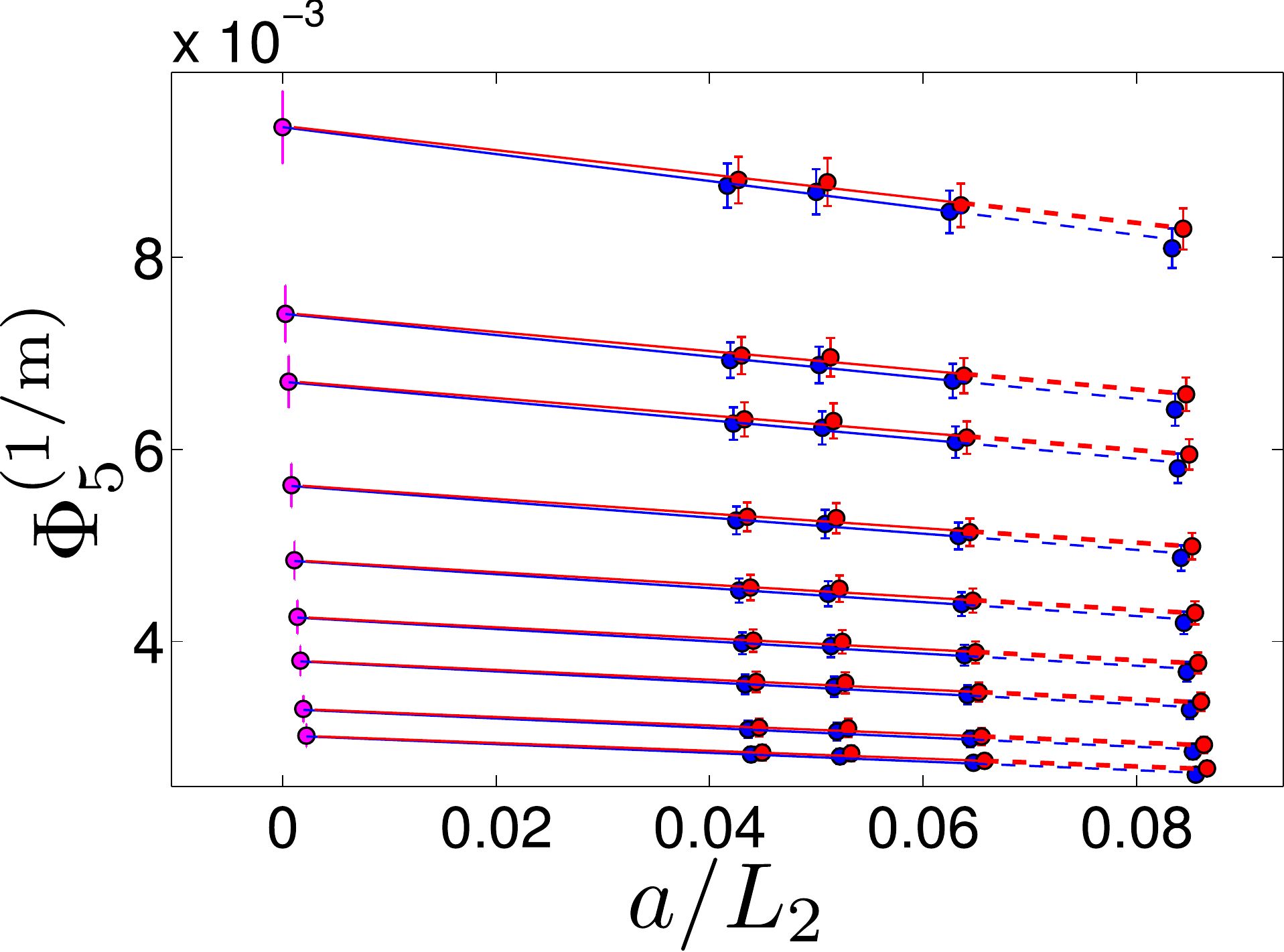}
\end{tabular}
\end{center}
\vspace{-0.5cm}
\caption[]{Same as Fig.~\ref{figPhi12_stat_L2} 
but for the $1/\mb$ terms $\Phi^\first_4$ and $\Phi^{\first}_5$,
 which are chosen to be proportional to $\omegakin$ and $\omegaspin$, respectively.}
\vspace{0.5cm}
\label{figPhi45_1m_L2}
\end{figure}

In general the equations~(\ref{eq:matchL1}) and (\ref{eq:matchL2})
are meant to be taken at the $1/\mb$ order of HQET, 
but they are of course valid in the static approximation
if one sets the various pieces to their static values.
In that case, the observables will be denoted by $\Phi^{\rm stat}$
, and otherwise we write $\Phi^{\rm HQET}$.
We also define the $1/\mb$ correction to be
$\Phi^{\first} = \Phi^{\rm HQET} - \Phi^{\rm stat} $.
As an illustration, we show some continuum extrapolations in the volume $L_2$:
the first two observables in the static approximation 
in \fig{figPhi12_stat_L2}, and the $1/\mb$ corrections $\Phi_4^{\first}$ and $\Phi_5^{\first}$
(which are sensitive to the kinetic and magnetic parameter $\omegakin$ and $\omegaspin$)
in  \fig{figPhi45_1m_L2}. 
Note that the $1/\mb$ terms are extrapolated linearly in the lattice spacing, whereas 
in the static case the continuum extrapolation is done quadratically in the lattice spacing.
This is justified because we use an $O(a)$-improved action 
and an $O(a)$-improved static axial current.
We end this section by collecting in Table~\ref{tabhqetparameters} the values of the HQET parameters
used in the numerical applications reported in the following.
\\

\section{Extraction of HQET hadronic matrix elements}

For the computation of the large-volume hadronic matrix elements,
we ought to take the continuum limit and extrapolate in the 
up/down quark mass to the physical point where $\mpi$ has its physical value.
The continuum extrapolation will be left for future work, we here just take
a quite small lattice spacing, $a=0.07\,\fm$. 
We note that in our quenched computations the difference between the continuum limit 
and $a=0.07\,\fm$ would be unnoticeable given the present errors.
Concerning the extrapolation in the light quark mass, we take into account 
terms at NLO in the chiral expansion for the static approximation, 
but neglect terms of order $m_\mrm{light} /m_\beauty$.

At the first order of the $1/\mb$ 
expansion our main observables are given by
\bes
\mB &=& m_{\rm bare} \,+\,E^{\rm stat}
\,+\, \omega_{\rm kin} \, E^{\rm kin} \,+\,
\omega_{\rm spin} \, E^{\rm spin}\,, 
\\
\mB-m_{\rm B^*} &=& {4\over 3} \omegaspin\,\Espin \;,
\\
\log(a^{3/2}\fB\sqrt{\mB/2}) &=& \log(\zahqet)+ \log(a^{3/2}p^{\rm stat})+ b^{{\rm stat}}_{\rm A} am_{\rm q} 
\nonumber \\&{\phantom{=}}& + \omega_{\rm kin}  p^{\rm kin} +
\omega_{\rm spin}  p^{\rm spin} + \cah{1} p^{\rm A^{(1)}} ,
\label{e:fbmb}
\ees
where $b^{{\rm stat}}_{\rm A}$ is an improvement coefficient (in practice we follow~\cite{Grimbach:2008uy} for its numerical
implementation).
The HQET energies and matrix elements have been measured on a subset of configuration ensembles 
produced within the CLS effort \cite{CLS} with 
${\rm N_{f}}=2$ flavors of 
${O}(a)$-improved Wilson-Clover fermions. We have collected in Table \ref{tabCLS} the main 
characteristics of that subset. 

\begin{table}
\begin{center}
\begin{tabular}{|c|c|c|c|c|c|c|c|c|c|}
\hline
$\beta$&$L\, M_Q$&$a\,m_{\rm bare}$&$\log(Z^{\rm HQET}_A)$&$[\log(Z_A)]^{\first}$ &$c^{(1)}_A/a$&$\omega_{\rm kin}/a$& $\omega_{\rm spin}/a$\\
\hline
$5.3$&$15$&$4 1.125(22)$&$-0.120(28)$& $0.039(26)$ &$-0.446(75)$&$0.476(06)$&$0.837(35)$\\
\hline
\end{tabular}
\end{center}
\caption{HQET parameters determined with the standard set of matching conditions,
that are used to obtain the central values of $m_{\rm b}$ and $f_{\rm B}$; the static quark action is HYP2
.} 
\label{tabhqetparameters}
\end{table}
\begin{table}[b]
\begin{center}
\begin{tabular}{cc}
\includegraphics*[width=1cm, height=0.5cm]{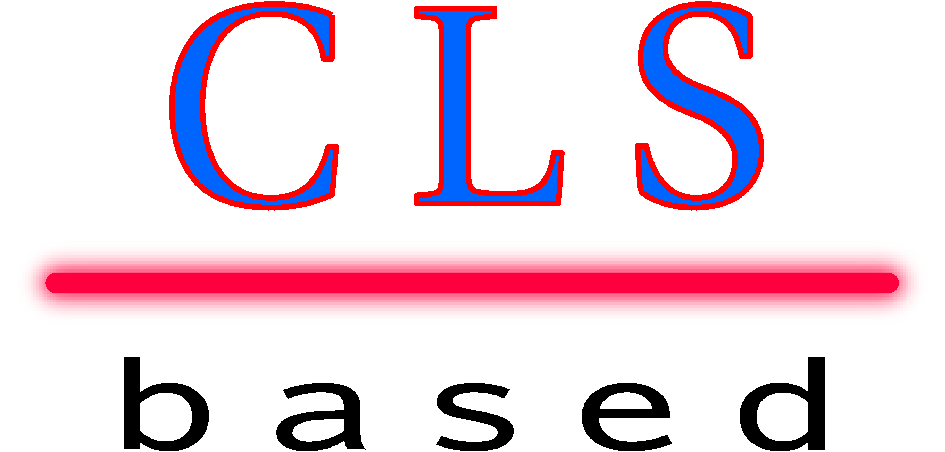}
&
\begin{tabular}{|c|c|c|c|c|c|c|}
\hline
$\beta$&$a$ (fm)&$L^{3}\times T$&$m_{\pi}$ (MeV)&\#&traj. sep.\\
\hline
5.3&0.07&$32^{3}\times 64$&550&152&32\\
&&$32^{3}\times 64$&400&600&32\\
&&$48^{3}\times 96$&300&192&16\\
&&$48^{3}\times 96$&250&350&16\\
\hline
\end{tabular}
\end{tabular}
\end{center}
\caption{Characteristics of the large volume simulations used so far to extract 
HQET energies and matrix elements. 
The last column is meant in terms of trajectories of length $\tau=1$.
}
\label{tabCLS}
\end{table}
\subsection{Large volume techniques}
\begin{figure}[b]
\begin{center}
\begin{tabular}{cc}
\includegraphics*[width=7.2cm]{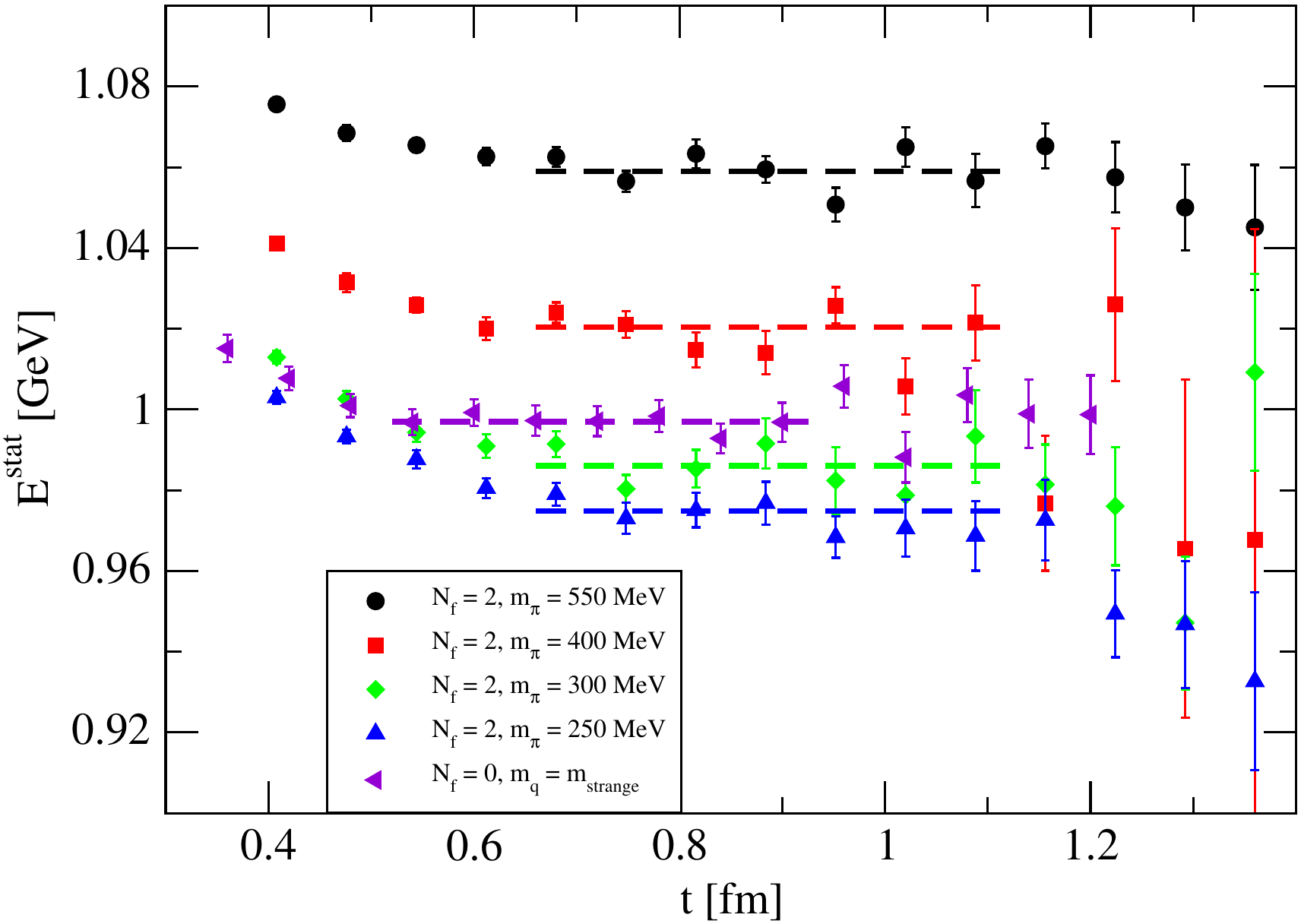}
&
\includegraphics*[width=7.2cm]{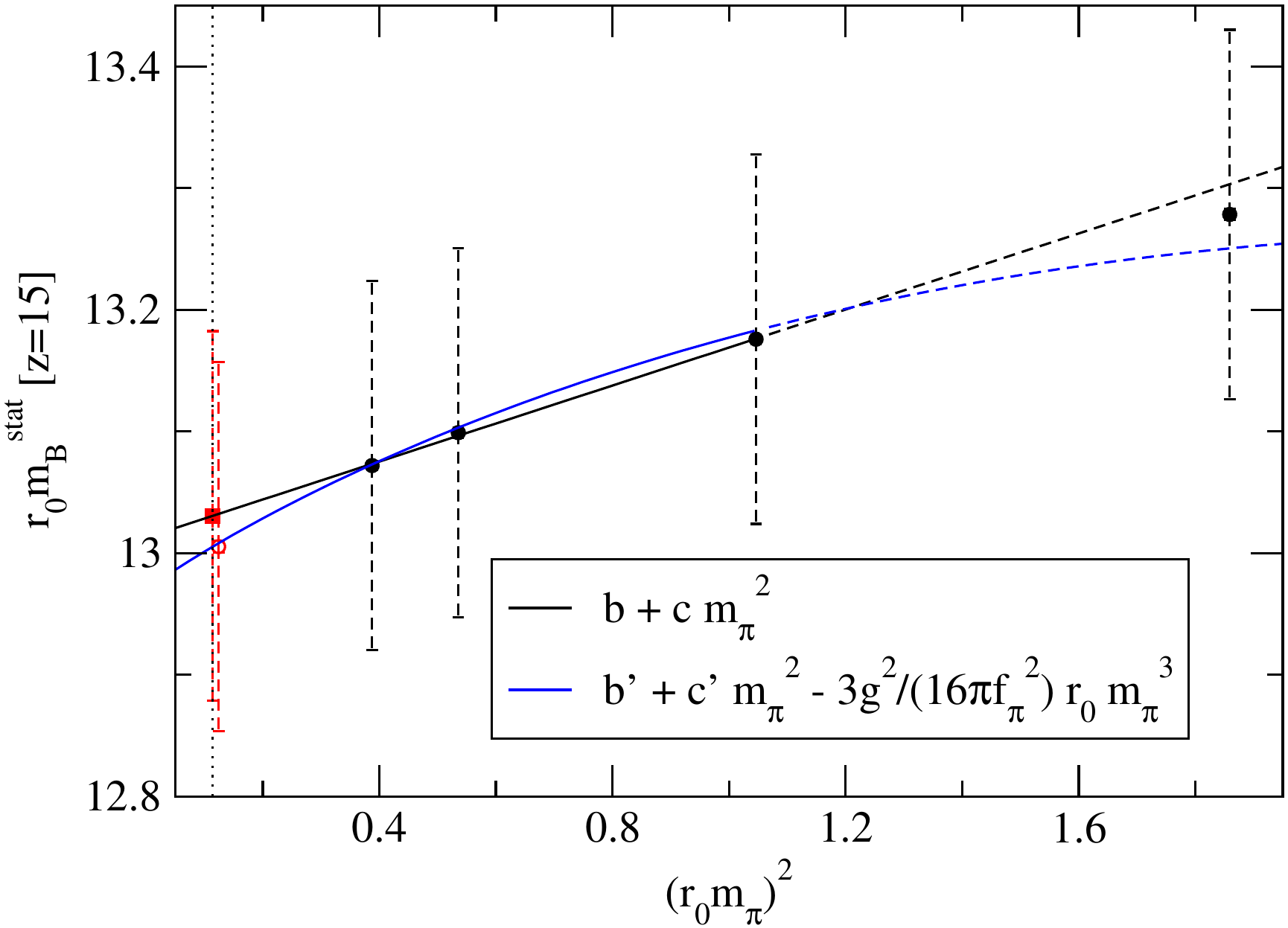}
\end{tabular}
\end{center}
\vspace{-0.5cm}
\caption{Plateaux of static energies with $N$=5 and $t_{0}=5a$ (left panel); 
the quenched data correspond to a lattice spacing quite close to the CLS set up at $\beta=5.3$. 
We remind the reader that the absolute value of $E^{\rm stat}$ has no real meaning without the 
subtraction of a linear $1/a$ divergence.
Right panel: chiral extrapolation of $r_{0} m^{\rm stat}_{\rm B}=r_{0}(E^{\rm stat} + m_{\rm bare}).$
A large part of the errors (dashed part), originating from $r_0$ and $\mhbare$,
is common to all data points and therefore irrelevant for the quark mass dependence. 
}
\vspace{0.5cm}
\label{figEstatnf0nf2Estatchiral}
\end{figure}

As interpolating fields we use quark bilinears
\begin{eqnarray}
O_k(x) = \psibar_{\rm h}(x)\gamma_0\gamma_5\psi_{\rm l}^{(k)}(x) \,, \quad
O_k^*(x) = \psibar_{\rm l}^{(k)}(x)\gamma_0\gamma_5\psi_{\rm h}(x) \,,
\end{eqnarray}
built from the static quark field $\psi_{\rm h}(x)$ and different levels
of Gaussian smearing \cite{wavef:wupp1} for the light quark field 
\begin{equation}
\psi_{\rm l}^{(k)}(x) = \left( 1+\kappa_{\rm G}\,a^2\,\Delta \right)^{R_k} \psi_{\rm l}(x) \,,
\end{equation}
with APE smeared links \cite{smear:ape,Basak:2005gi} in the lattice Laplacian $\Delta$, and
with the same parameters as in \cite{BlossierVZ}. We compute the following correlators
\bea
C^{\rm{stat}}_{ij}(t)& = &
\sum_{x,\bf y}
\left< O_i(x_0+t,{\bf y})\,O^*_j(x)\right>_\stat,\\
C^{\rm{kin/spin}}_{ij}(t)& = &
\sum_{x,{\bf y},z}
\left< O_i(x_0+t,{\bf y})\,O^*_j(x)\,  O_{\rm kin/spin}(z)\right>_\stat \,,
\\
C^{\rm{stat}}_{A^{(1)},i}(t)& = &
\sum_{x,\bf y}
\left< A^{(1)}_0(x_0+t,{\bf y}) O^*_i(x)\right>_\stat \,,
\eea
using stochastic all-to-all propagators to reduce the variance.
From the $N\times N$ matrices 
of correlators $C^{\rm stat}$, $C^{\rm kin/spin}$, and $C^{\rm stat}_{A^{(1)}}$, we solve the generalised 
eigenvalue problem (GEVP)
\begin{equation}
C(t)v_{n}(t,t_{0})=\lambda_{n}(t,t_{0}) C(t_{0}) v_{n}(t,t_{0})\,,
\end{equation}
to get the energies and operators having the largest overlap with the $n^{\rm th}$ state:
\bea
aE^{\rm eff}_{n}(t,t_{0})&=& 
-\ln\left(\frac{\lambda_{n}(t+a,t_{0})}{\lambda_{n}(t,t_{0})}\right)\;,\\
Q_{n}^{\rm eff}(t,t_{0})&=& 
\frac{O^{i}(t) v^i_n(t,t_{0})}
{\sqrt{v^i_n(t,t_{0})C_{ij}(t) v^j_n(t,t_{0})}}
\left(\frac{\lambda_{n}(t_{0}+a,t_{0})}{\lambda_{n}(t_{0}+2a,t_{0})}\right)^{t/2a}\;.
\eea
For a computation of the $1/\mb$ corrections to static 
energies and matrix elements one needs to solve the GEVP at the static order 
only\cite{BlossierKD}. For instance, in the case of 
energies, one has
\bea
E^{\rm eff}_{n}(t,t_{0})&=&E^{\rm eff, stat}_{n}(t,t_{0}) + \omega E^{\rm eff, 1/m}_{n}(t,t_{0})
+ O(\omega^2)\;,\\
aE^{\rm eff, stat}_{n}(t,t_{0})&=& 
-\ln\left(\frac{\lambda^{\rm stat}_{n}(t+a,t_{0})}{\lambda^{\rm stat}_{n}(t,t_{0})}\right)\;,\\
E^{\rm eff, 1/m}_{n}(t,t_{0})&=& 
\frac{\lambda^{\rm 1/m}_{n}(t,t_{0})}{\lambda^{\rm stat}_{n}(t,t_{0})}
-\frac{\lambda^{\rm 1/m}_{n}(t+a,t_{0})}{\lambda^{\rm stat}_{n}(t+a,t_{0})}\;,\\
\frac{\lambda^{\rm 1/m}_{n}(t,t_{0})}{\lambda^{\rm stat}_{n}(t,t_{0})}&
=&\sum_{i,j} v^{\rm stat}_{n\,i}(t,t_0) \left[\frac{C^{\rm 1/m}_{ij}(t)}{\lambda^{\rm stat}_n(t,t_0)}
-C^{\rm 1/m}_{ij}(t_0)\right] v^{\rm stat}_{n\,j}(t,t_0)\;.
\eea
As illustrated in the left panel of \fig{figEstatnf0nf2Estatchiral}, a
comparison with quenched data indicates that the statistical error on static energies is not an issue, 
even at small quark masses.
Some care has to be taken to control the contribution of excited states to 
extracted energies. For the energy levels determined from the GEVP, 
the leading corrections
are given by
\begin{equation}
\label{e:correction}
E^{\rm eff, stat}_n(t,t_0) = E^{\rm stat}_n + \beta^{\rm stat}_n
e^{-\Delta E_{N+1,n}t} + \ldots
\end{equation}
where $\Delta E_{m,n}=E_{m}-E_{n}$ and the condition $t_0\geq t/2$ is necessary 
to prove \eq{e:correction}. In our analysis we have chosen a time range to extract the 
plateaux such that the corrections to $E^{\rm stat}_{1}$ are small compared to its statistical error; 
we found this to be the case for $t_{0}>0.3$ fm.

\begin{figure}[t]
\begin{center}
\begin{tabular}{cc}
\includegraphics*[width=7.2cm]{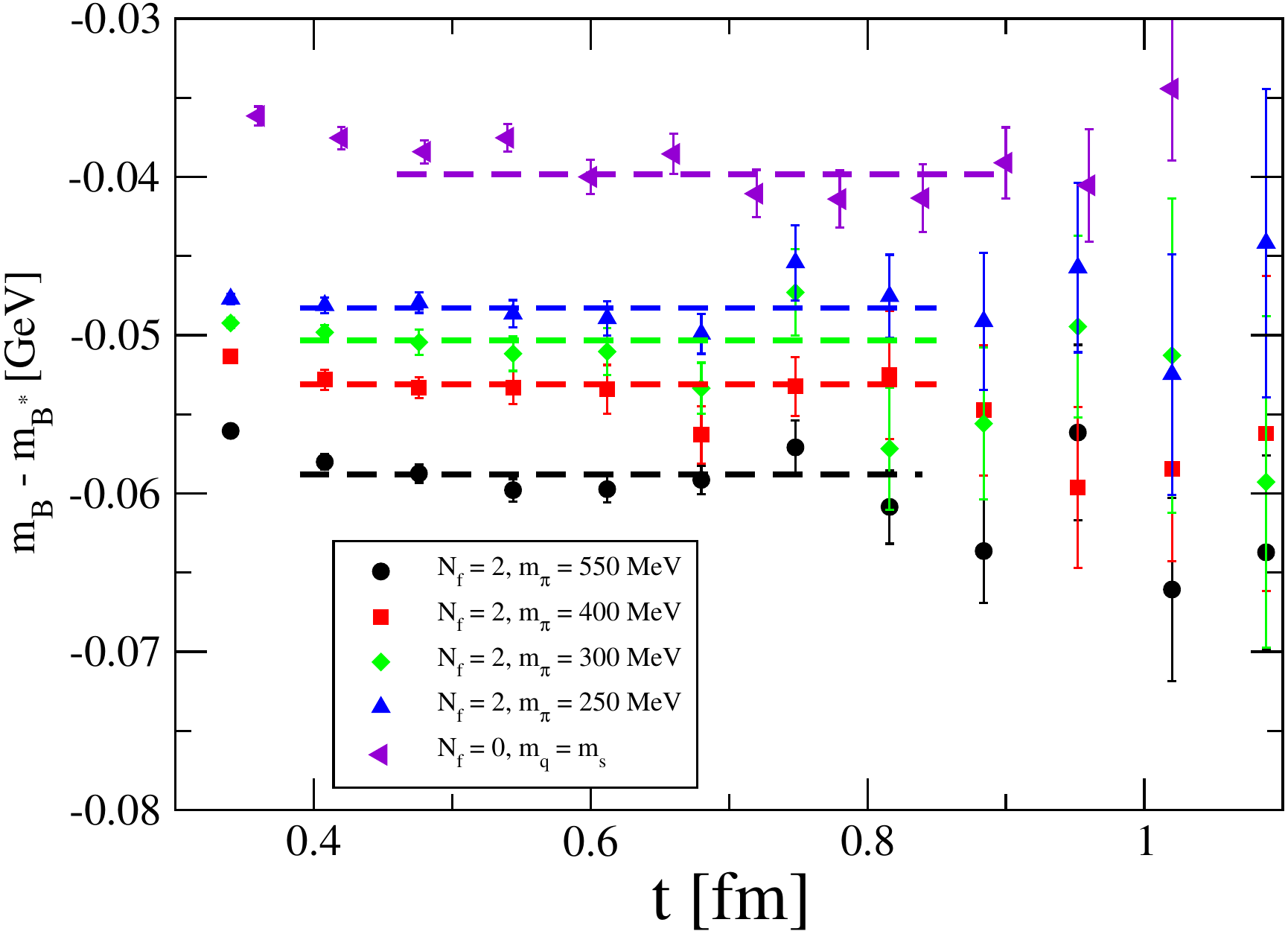}
&
\includegraphics*[width=7.2cm]{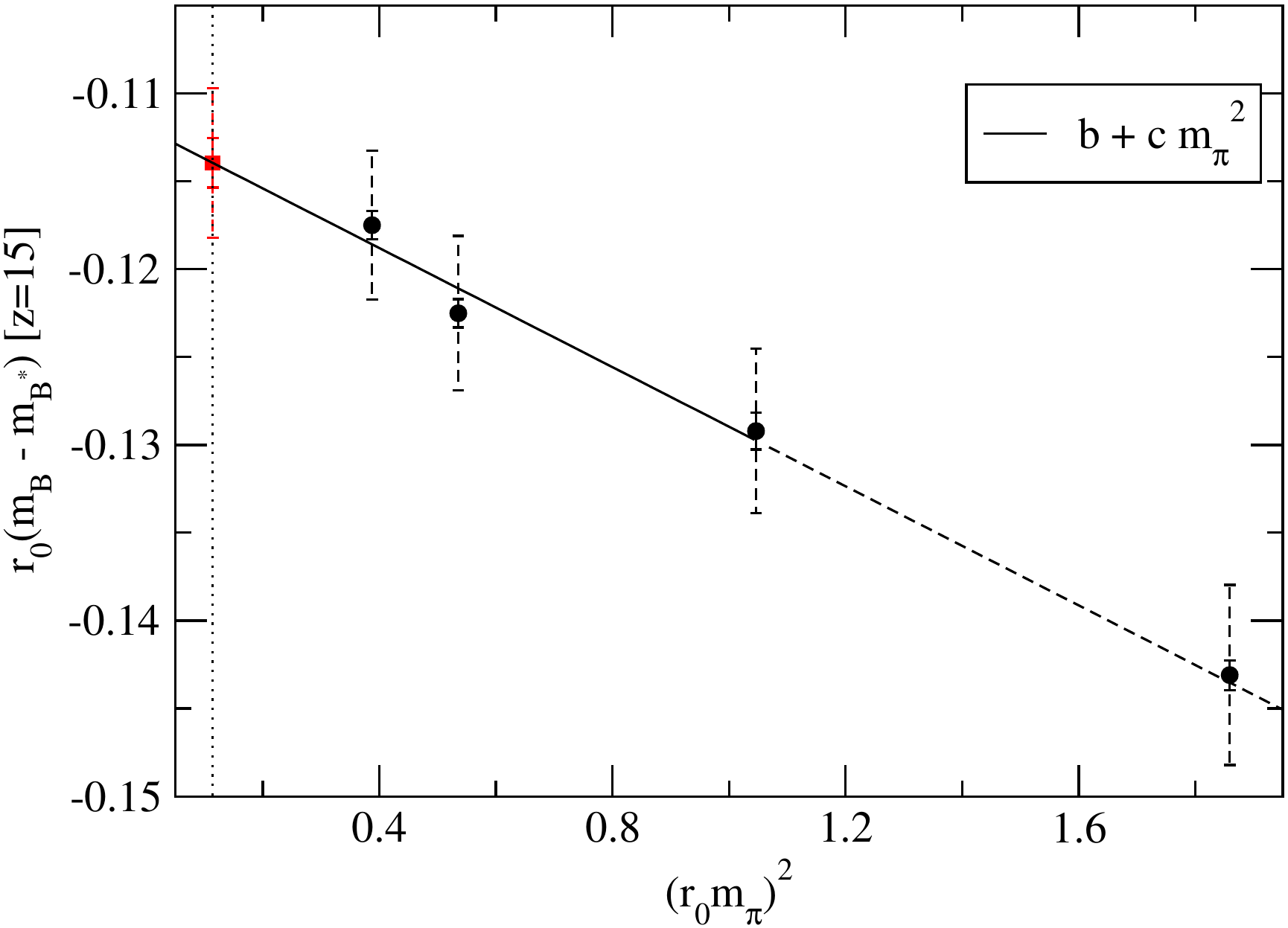}
\end{tabular}
\end{center}
\vspace{-0.5cm}
\caption{On the left panel we show the plateaux of $\omega_{\rm spin}E^{\rm spin}$ with $N$=5 and $t_{0}=4a$; 
the quenched data correspond to a lattice spacing quite close to the CLS set up at $\beta=5.3$. 
On the right panel we show the chiral extrapolation of $r_{0} (\mB-m_{\rm B^*})$ with a LO fit.
Here the dashed error due
to $\omegaspin$ and $r_0$ is independent of the quark mass.
}
\vspace{0.5cm}
\label{figEspin}
\end{figure}
\subsection{$b$-quark mass}
Once the energies are obtained at the simulated dynamical
quark masses, we still need to extrapolate them to the physical point. 
We have chosen pion masses below $450$ MeV for the extrapolation 
of all quantities reported in these proceedings. For the extrapolation
of $m^{\rm stat}_{\rm B} \equiv E^{\rm stat}+m^{\rm stat}_{\rm bare}$
we have used the following form 
\bea
\label{e:chirfit}
r_{0} \mB^{\rm stat}&=& b+ c m^{2}_{\pi} + d m^{3}_{\pi}\,,
\eea
where we once set $d=0$, and once use 
$d=- 3 r_0 \hat{g}^2 / (16 \pi f^2_\pi) $ computed with 
the experimental value of $f_\pi$ and the recent lattice determination of
$\hat{g}=0.51(2)$ \cite{MDonnellan}.
We show these fits in the right panel of 
\fig{figEstatnf0nf2Estatchiral}. 
As our central value of $r_{0}\mB^{\rm stat}$ at the physical point we 
take the average of the two extrapolations. Half the difference
is included in the systematic errors.
At present a much larger source of uncertainty is
$r_{0}/a$, whose determination is discussed in \cite{leder}. Moreover, an 
uncertainty of 5\% has been added on the scale $r_{0}$ itself. 
We take $r_{0}=0.475 \pm 0.025$ fm \cite{Etm2010,milc:r0,bmw:r0}. Identifying 
the static approximation result, $\mB^{\rm stat}$, with $\mB^\mrm{exp}$, 
we can determine the RGI $b$-quark mass from static HQET.
Translating as usual to the $\overline{\rm MS}$ scheme with 4-loop perturbation theory
and the known $\Lambda$-parameter \cite{alpha:nf2}, we find

\begin{equation}
\mb^{\overline{\rm MS}}(\mb^{\overline{\rm MS}})^{\rm stat}_{\rm N_f=2}=4.255(25)_{r_0}(50)_{\rm{stat+renorm}}(?)_a\;\;
\mbox{GeV}\,,
\end{equation}
where the first error comes from the uncertainty on $r_0$, while the second error includes the statistical 
error on $aE^{\rm stat}$, the uncertainty on the chiral extrapolation and the error on the
quark mass renormalisation constant $Z_{\rm M}$ in QCD \cite{heit_frit_tant}. 
The latter is currently dominating. The ``$(?)_a$'' indicates that a continuum 
limit is not yet performed, but as mentioned earlier we expect only a smaller 
error due to that\cite{BlossierJK}.

Amongst the $O(1/\mb)$ corrections to energies the 
hyperfine mass splitting is particularly interesting since here the $1/\mb$ term is 
the dominant one and is given by a single HQET parameter, $\omegaspin$. 
This contribution can hence be discussed separately. Its determination is encouraging, as far as the statistical
uncertainty and the chiral extrapolation are concerned, as shown in \fig{figEspin}. 
Plateaux have the same quality as in the quenched case and the statistical precision 
is good enough to perform a reasonable chiral extrapolation. Since we have seen in the 
quenched approximation that cut-off effects can be sizable for this quantity, we
do not quote the hyperfine splitting in MeV at the moment.

Including the magnetic as well as the kinetic correction, we compute $\mB^{\rm HQET}\equiv\mB^{{\rm stat+1/m}}$,
perform the chiral extrapolation as for $\mB^{\rm stat}$, and from equating $\mB^{\rm HQET}=\mB^{{\rm exp}}$
we obtain the $b$-quark mass up to tiny $\Lambda^3/\mbeauty^2$ effects as
\begin{equation}
\left.\mb^{\overline{\rm MS}}(\mb^{\overline{\rm MS}})\right\vert^{{\rm HQET}}_{{\rm N_f}=2}=
4.276(25)_{r_0}(50)_{\rm{stat+renorm}}(?)_a\;\;
\mbox{GeV}\,.
\end{equation}
Sources of errors are the same as in the static case. Some previous determinations read\
\begin{center}
\begin{tabular}{rcl}
$\left.\mb^{\overline{\rm MS}}(\mb^{\overline{\rm MS}})\right\vert^{{\rm HQET}}_{\rm N_{\rm f}=0}$ & 
$=$ & 
$4.320(40)_{r_{0}}(48)\;\;\mbox{GeV}\quad$
\cite{DellaMorteCB}\,,
\\
$\mb^{\overline{\rm MS}}(\mb^{\overline{\rm MS}})$&
$=$&
$4.163(16)\;\; \mbox{GeV}\quad$
\cite{ChetyrkinFV}\,.
\end{tabular}
\end{center}

In the future, a nice check of our result
will consist in doing the same computation in a partially quenched set up; 
there, the experimental inputs will be the
$B_{\rm s}$ spectrum, and the determination of the hopping parameter 
of the strange quark $\kappa_{\rm s}$ will be necessary. 
We will then be able to directly observe possible quenching effects 
because in the comparison with
\cite{DellaMorteCB} we can use exactly the same experimental input. 
While the agreement with 
\cite{ChetyrkinFV} is not convincing at the moment, it does not seem
worrying given the present errors. We emphasize that within 
about 100 MeV all these numbers agree, while a more precise statement needs
additional work, in particular the continuum limit on our side, and on a longer
term $\rm N_f>2$.

\subsection{$B$ meson decay constant and $V_{\rm ub}$}
\begin{figure}[t]
\begin{center}
\begin{tabular}{cc}
\includegraphics*[width=7.2cm]{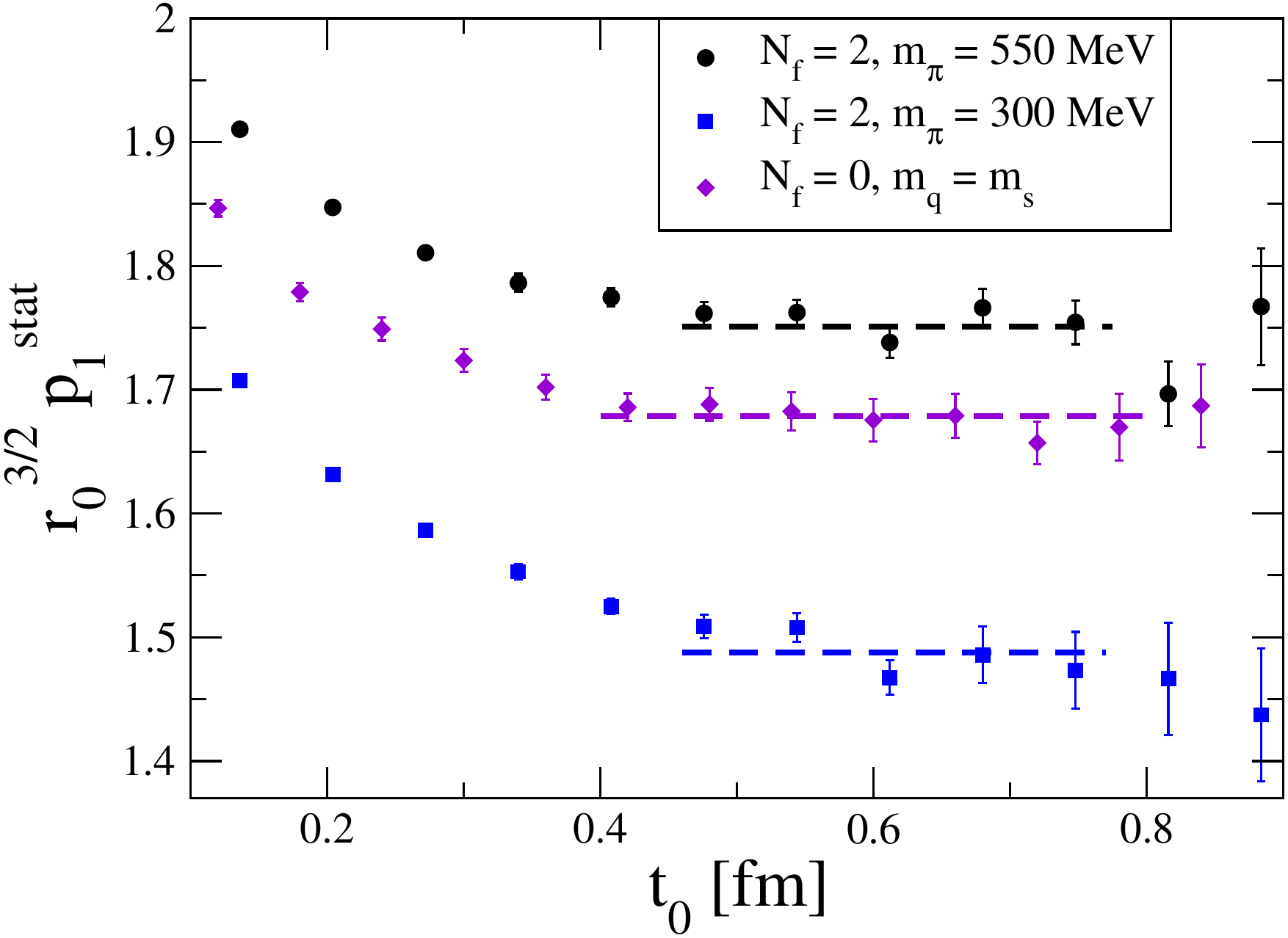}
&
\includegraphics*[width=7.2cm]{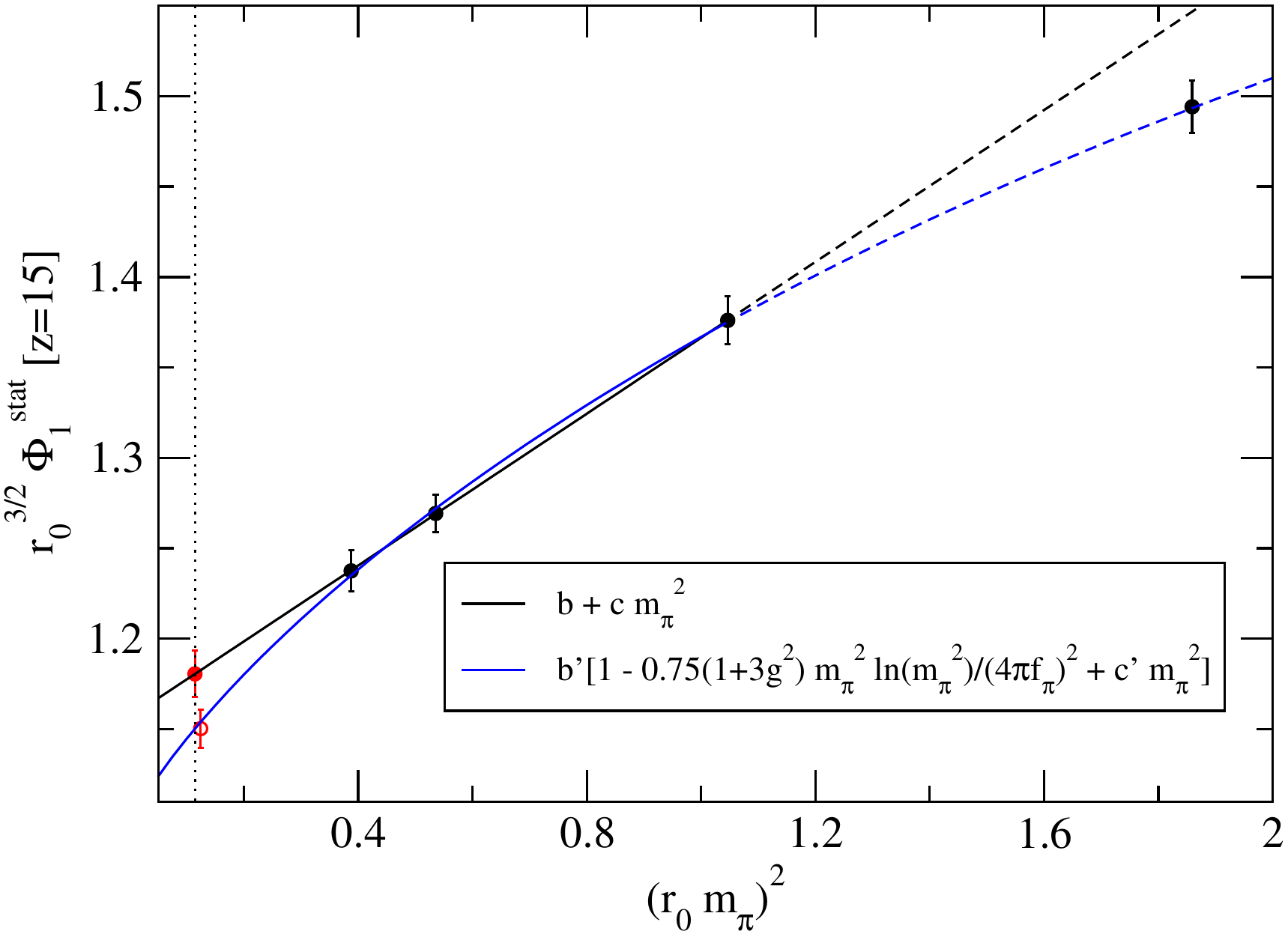}
\end{tabular}
\end{center}
\vspace{-0.5cm}
\caption{On the left panel we show the plateaux of $r^{3/2}_0 p^{\rm stat}_1$ with $N$=3 and $t-t_{0}\sim 0.3$ fm; 
the quenched data correspond to a lattice spacing quite close to the CLS setup at $\beta=5.3$. 
On the right panel we show the chiral extrapolation of $r^{3/2}_{0} \Phi^{\rm stat}_1$. }
\vspace{0.5cm}
\label{figfB}
\end{figure}

\begin{figure}[b]
\begin{center}
\begin{tabular}{cc}
\includegraphics*[width=7.0cm]{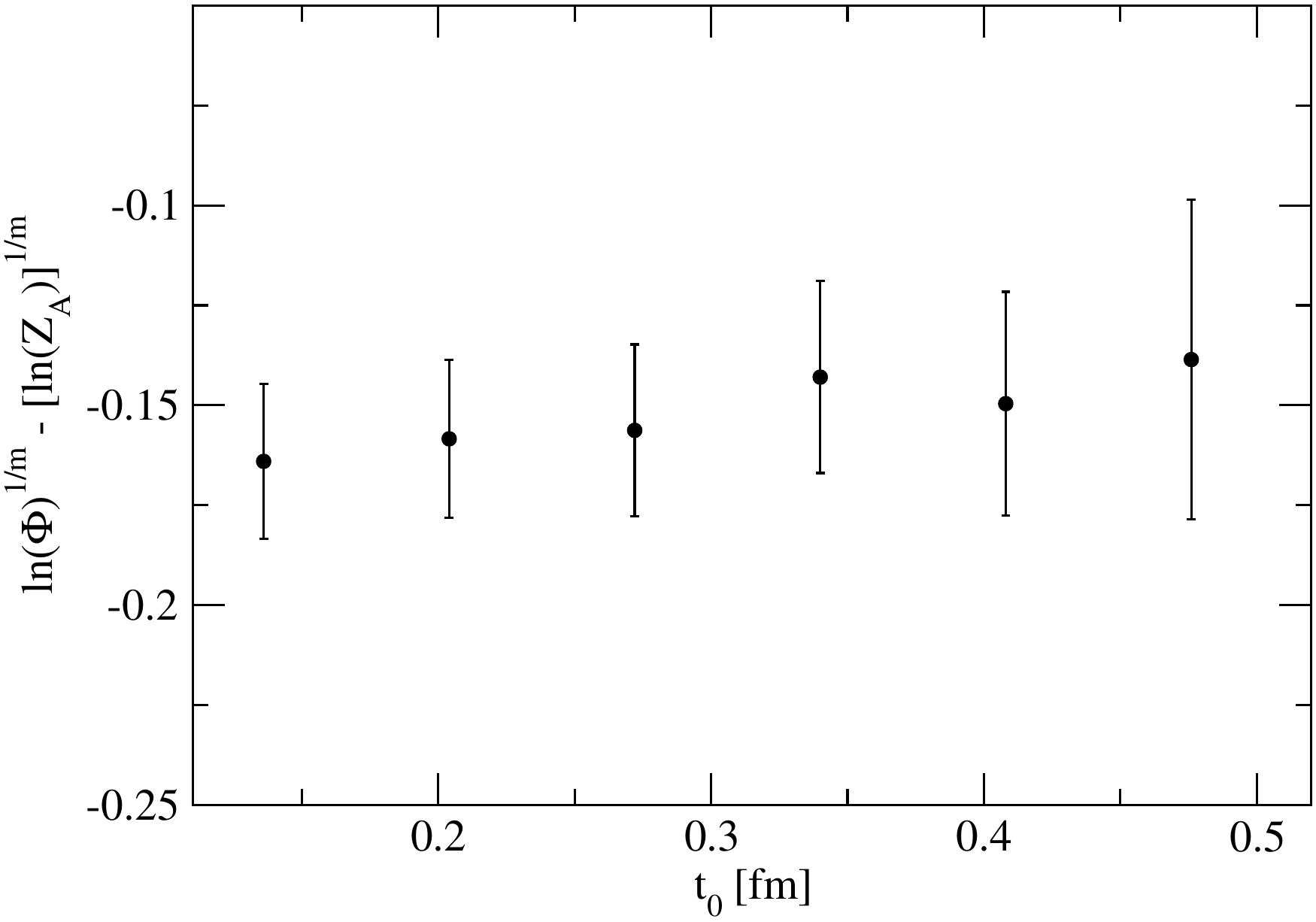}
&
\includegraphics*[width=7.0cm]{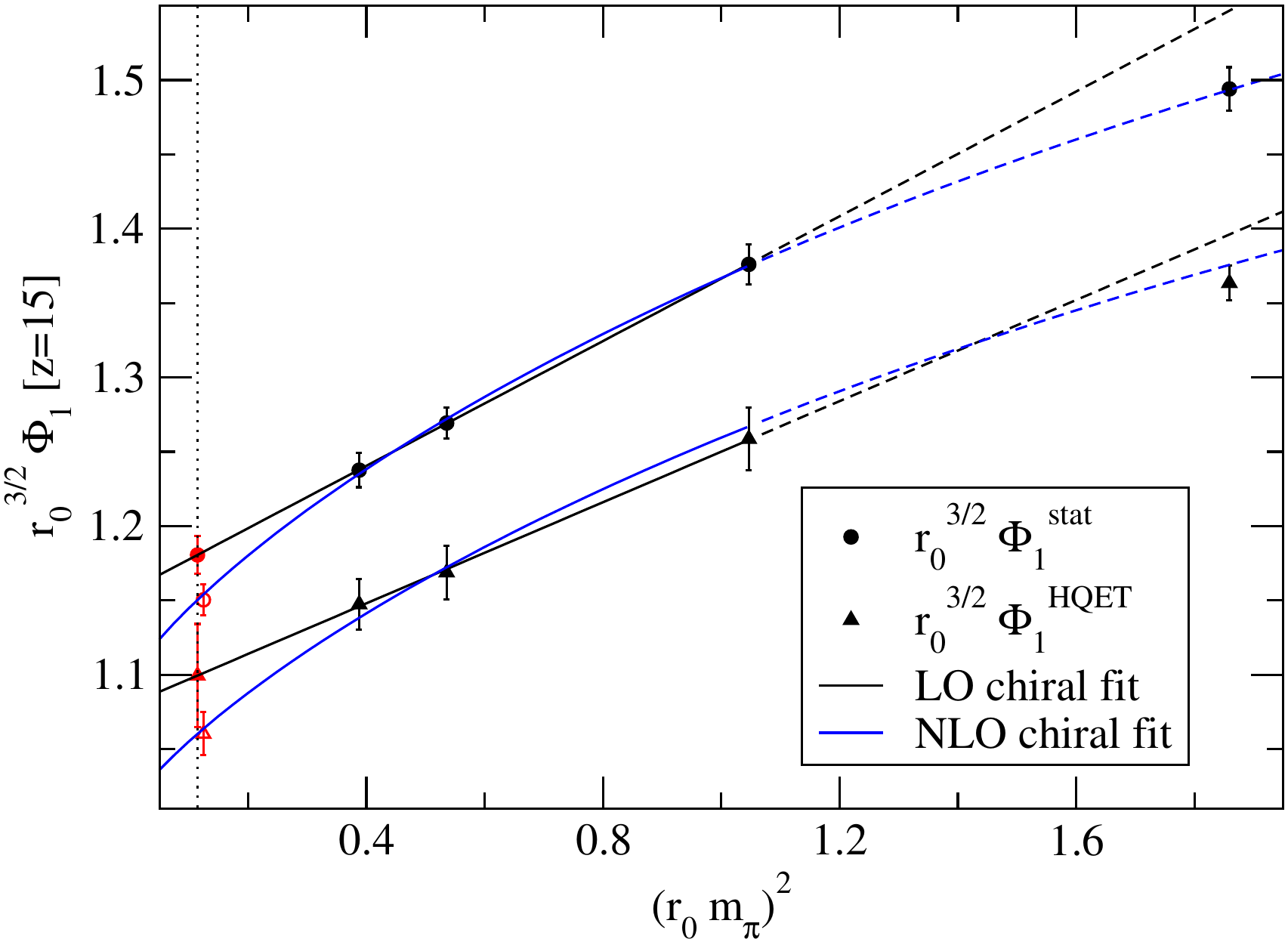}
\end{tabular}
\end{center}
\vspace{-0.5cm}
\caption{Plateau of $\log(\Phi_1)^{\first}-[\log(Z_A)]^{\first}$ for $N$=4,  $t-t_{0}=4a$, $z=15$ and $m_\pi \simeq 550$ MeV
(left panel) and chiral extrapolation at the static order and at order $1/m$ (right panel).}
\vspace{0.5cm}
\label{figfB1om}
\end{figure}

Let us now discuss  $\fB$. In HQET we consider the particular combination~\footnote{
We warn the reader that in this section we follow the notations introduced 
in~\cite{BlossierMK}, where the subscript 1 represents the ground state. 
In particular $\Phi_1$ should not be confused with the finite volume meson
mass $\Phi_1$ introduced in section~2.}
\begin{equation}
\Phi_1 \;\equiv\; \fB\,\sqrt{\mB/2} \,,
\end{equation} 
see \eq{e:fbmb}. We separate the static and the $1/{m_{\rm b}}$ contributions as
\begin{eqnarray}
\log(r_0^{3/2}\Phi_1) &=& \log(r_0^{3/2}\Phi_1^{\rm stat}) + [\log(\Phi_1)]^{\first} \;, \\{}
[\log(\Phi_1)]^{\first}&=& [\log(Z_A)]^{\first} + 
\omega_{\rm kin}  p^{\rm kin} +
\omega_{\rm spin}  p^{\rm spin} + (\cah{1}-a\castat) p^{\rm A^{(1)}} .
\end{eqnarray}
On the left panel of \fig{figfB} we show that the GEVP method works 
as well as in the quenched case to extract the static matrix element. 
A good plateau is visible and our confidence in it is also based on 
the knowledge that corrections are $O(e^{-\Delta E_{N+1,n}t_0})$ when the
computation is done as here (see~\cite{BlossierKD,BlossierMK}).
We have again applied two kinds of 
extrapolations of the static approximation
$r^{3/2}_0 \Phi^{\rm stat}_1$ to the physical point \cite{SharpeQP}:
\begin{eqnarray}
r^{3/2}_0 \Phi^{\rm stat}_1 &=& b+c m^2_\pi \quad (\mbox{LO})\,,\\
r^{3/2}_0 \Phi^{\rm stat}_1 &=& b'\left[1-\frac{3}{4} \frac{1+3\hat{g}^2}{(4\pi f_\pi)^2}
m^2_\pi\ln (m^2_\pi)+c' m^2_\pi\right] \quad (\mbox{HM}\chi\mbox{PT})\,.
\end{eqnarray}
Again we fixed $\hat{g}$ \cite{MDonnellan}. The same fit form is used for
$r_0^{3/2} \Phi_1^{\rm HQET}$.
As shown in the right panels of \fig{figfB} and \fig{figfB1om}, whether we do 
or do not
include the chiral logarithm of 
HM$\chi$PT, changes the value at the physical point by a small
but noticeable amount. 
At the moment we take the average of the two extrapolations
as the central value and include half of the difference as part of the systematic error. We obtain
\be
 {\fB}^{\rm HQET}_{\; \rm N_{\rm f}=2}=178(16)(?)_{a}\,\MeV\,,  
\ee
where the first
error includes the statistical uncertainty on matrix elements, 
the systematics coming from chiral extrapolation and the uncertainty on the
physical scale of $r_{0}$, while cut-off effects are not estimated yet. 
Let us recall that in the quenched approximation we have 
obtained ${\fBs}^{\rm HQET}_{\; \rm N_{\rm f}=0}=234(18)$ MeV~\cite{PapIII}
for $r_0=0.475$ fm and the error covers for a 5\% uncertainty on $r_0$. 

\begin{figure}
\begin{center}
\includegraphics*[width=12cm, height=8cm]{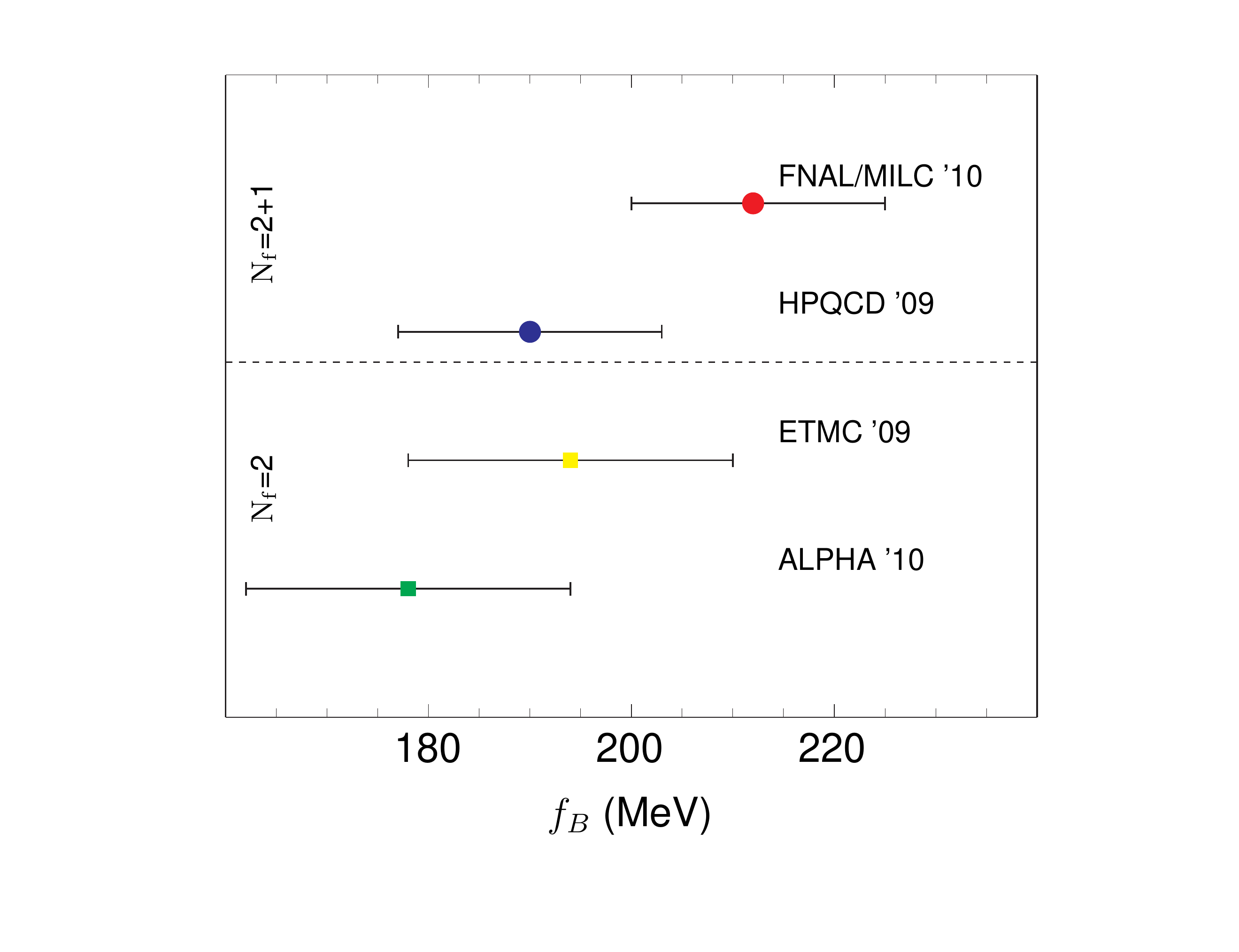}
\end{center}
\vspace{-1cm}
\caption{Collection of recent lattice computations of $f_{\rm B}$.
\label{figfBlattice}}
\end{figure}

At present there is a tension between the two ways of determining 
$V_{\rm ub}$ from the exclusive $B$-decays $B \to \pi l \nu$
and $B \to \tau \nu$ \cite{Porter}. Both of them rely on the determination 
of the hadronic matrix elements from lattice QCD. In the Standard Model, the branching ratio
for the latter decay is simply
proportional to $f_{\rm B}^2 \vert V_{\rm ub}\vert^2$ with all other factors known. 
In \fig{figfBlattice}, we have collected the 
values of $f_{\rm B}$ obtained recently by several collaborations \cite{resultsfBlat}, 
from ${\rm N_{f}=2}$ (ALPHA, ETMC) and ${\rm N_{f}=2+1}$ (FNAL/MILC, HPQCD) simulations. 
The heavy quark is discretised differently in each of the 
quoted computations: NRQCD (HPQCD), Fermilab action (FNAL/MILC), 
Twisted-Mass QCD (ETMC) and HQET (ALPHA). 
The systematic errors and difficulties in these different approaches
are therefore rather different. For NRQCD a continuum limit does not exist
and one studies a range of lattice spacings in the window between sizable
discretisation errors and the divergent behaviour. In the relativistic 
twisted-mass approach one extrapolates in the heavy quark mass by assuming a HQET inspired
form, and a continuum extrapolation has been performed. 
Some computations rely on perturbatively determined parameters.
The FNAL/MILC and
HPQCD computations use 1-loop renormalization\footnote{\label{NPoneloop}FNAL/MILC splits the
renormalization factor into a non-perturbatively computed part and a rest estimated
by 1-loop perturbation theory.} for the currents and part of the parameters in the 
Lagrangian. In our approach, all renormalization
is done non-perturbatively and the theory has a continuum limit, but so far only
a single lattice spacing of $a=0.07$ fm is available in large volume. There is an intrinsic
truncation error of $\rmO(\Lambda^2/m^2)$ which is present (more or less explicitly) 
in all approaches.\footnote{Generically the truncation error is $\Lambda^2/\mbeauty^2$, but for
the extrapolation in $m$ significantly smaller masses enter. }
We have sketched in Table \ref{tabgroups} some of these issues. Despite the rather 
different characteristics, the results are very similar. The range of $f_{\rm B}$ 
in \fig{figfBlattice} does not reach large enough values
to remove the tension between $B \to \tau \nu$ and $B \to \pi l \nu$ branching ratios 
within the Standard Model, and from \fig{figfBlattice} it seems very unlikely that 
the lattice determinations of $\fB$ are the reason for the tension.
However, interpreting this as a hint
for physics beyond the Standard Model, one has to keep in mind
that the $B \to \pi l \nu$ 
form factors are less well studied than $f_{\rm B}$ and that the experimental
determination of the branching ratio $B \to \tau \nu$ is rather difficult. 

\begin{table}
\begin{center}
\begin{tabular}{|c|c|c|c|}
\hline
Group & Method                 & Renormalisation & Range in $a$ \\
\hline
ALPHA & HQET at $O(1/\mb)$    & non-perturbative & $0.07\,\fm$  \\
ETMC  & twisted mass           &  non-perturbative &  $[0.065\,-\,0.1]\, \fm$            \\[-1.0ex]
      & extrapolation in $\mb$ &                 &              \\ 
HPQCD & NRQCD                  & 1-loop PT       &  $[0.09\,-\,0.12]\, \fm$           \\
FNAL/MILC & Fermilab action    & 1-loop PT$^{3} 
                                           $     &  $[0.09\,-\,0.12]\, \fm$            \\
\hline
\end{tabular}
\end{center}
\caption{Present methods in lattice computations of $f_{\rm B}$ in unquenched
simulations. \label{tabgroups}}
\end{table}

\section{Conclusions}

We have reported on the status of the project undertaken by the ALPHA Collaboration to extract 
relevant $B$ physics quantities 
from ${\rm N_{\rm f}}=2$ lattice simulations 
in the framework of HQET expanded at $O(1/\mb)$. 
The non-perturbative matching of HQET with QCD, through simulations performed in a small volume
$L_1\sim$ 0.5 fm, is almost done. 
The measurement of HQET energies and matrix elements, using the GEVP approach, has 
started recently on ensembles produced by CLS. The analysis of quantities like 
the $b$-quark mass and the $B$ decay constant is on the way.
The first results are promising and once we have controlled cut-off effects 
by simulation at several lattice spacings, we will also determine
hadronic parameters of the $B-\bar{B}$ mixing or $B \to \pi$ semileptonic form factors
as well as more details of the spectrum of hadrons with a b-flavor.

\acknowledgments
Work supported in part by the SFB/TR~9  and grant HE~4517/2-1 of the Deutsche Forschungsgemeinschaft
and by the European Community
through EU Contract No.~MRTN-CT-2006-035482, ``FLAVIAnet''. 
We thank F. Bernardoni for discussions on HMChPT, and we thank
CLS for the joint production and use of gauge configurations\cite{CLS}. 
Our simulations are performed on BlueGene, PC-clusters, and apeNEXT 
of the John von Neumann Institute for Computing at FZ J\"ulich, of the HLRN 
in Berlin, and at DESY, Zeuthen.
We thankfully acknowledge the computer resources and support provided by
these institutions.

\end{document}